\documentclass[conference]{llncs}

\usepackage{amsfonts}
\usepackage{amssymb}
\usepackage{stfloats}
\usepackage{cite}
\usepackage{graphicx}
\usepackage{psfrag}
\usepackage{subfigure}
\usepackage{amsmath}
\usepackage{array}
\usepackage{algorithm}
\usepackage{algpseudocode}
\usepackage[affil-it]{authblk}
\usepackage[english]{babel}
\usepackage{blindtext}
\usepackage{url}
\usepackage{epstopdf}
\usepackage{flushend}
\usepackage{multirow}

\title{Deep Convolutional Neural Network with Mixup for Environmental Sound Classification}

\author{Zhichao Zhang\and Shugong Xu\thanks{Corresponding author.
Shanghai Institute for Advanced Communication and Data Science,
Shanghai University, Shanghai, China(email: \email{shugong@shu.edu.cn}).}\and Shan Cao\and Shunqing Zhang}
\institute{Shanghai Institute for Advanced Communication and Data Science,\\
Shanghai University, Shanghai, 200444, China\\
\email{\{zhichaozhang, shugong, cshan, shunqing\}@shu.edu.cn}}

\begin{document}

\maketitle

\begin{abstract}

Environmental sound classification (ESC) is an important and challenging problem. In contrast to speech, sound events have noise-like nature and may be produced by a wide variety of sources.  In this paper, we propose to use a novel deep convolutional neural network for ESC tasks. Our network architecture uses stacked convolutional and pooling layers to extract high-level feature representations from spectrogram-like features. Furthermore, we apply mixup to ESC tasks and explore its impacts on classification performance and feature distribution. Experiments were conducted on UrbanSound8K, ESC-50 and ESC-10 datasets. Our experimental results demonstrated that our ESC system has achieved the state-of-the-art performance (83.7$\%$) on UrbanSound8K and competitive performance on ESC-50 and ESC-10. 


\end{abstract}


\section{Introduction}
Sound recognition is a front and center topic in today's pattern recognition theories, which covers a rich variety of fields. Some of sound recognition topics have made remarkable research progress, such as automatic speech recognition (ASR)\cite{hinton2012deep,graves2013speech} and music information retrieval (MIR)\cite{casey2008content,typke2005survey}. 
Environmental sound classification (ESC) is an another important branch of sound recognition and is widely applied in surveillance\cite{Radhakrishnan2005Audio}, home automation\cite{Vacher2014Sound}, scene analysis\cite{Barchiesi2015Acoustic} and machine hearing\cite{Lyon2010Machine}. However, unlike speech and music, sound events are more diverse with a wide range of frequencies and often less well defined, which make ESC tasks more difficult than ASR and MIR. Hence, ESC still faces critical design issues in performance and accuracy improvement.

Traditional ASR techniques such as MFCC, LPC, PLP are applied directly to ESC fields in previous works\cite{eronen2006audio,temko2006comparison,lee2010audio,mcloughlin2008line}. However, state-of-the-art performance has been achieved when using more discriminative representations such as Mel filterbank features\cite{Chu2009Environmental}, Gammatone features\cite{Valero2012Gammatone} and wavelet-based features\cite{Geiger2015Improving}. These features were modeled with some typical machine learning algorithms such as SVM\cite{uzkent2012non}, GMM\cite{mesaros2018detection} and KNN\cite{Piczak2015ESC} for ESC tasks. However, the performance gain introduced by these approaches is still unsatisfying. One main reason is that traditional classifiers do not have feature extraction ability.

Over the past few years, deep neural networks (DNNs) have made great success in ASR and MIR\cite{hinton2012deep,schedl2014music}. For audio signals, DNNs have ability to extract features from raw data or hand-draft feature. Therefore, some DNN-based ESC systems\cite{mcloughlin2015robust,kons2013audio} were proposed and performed much better than SVM-based ESC system. However, deep fully-connected architecture of DNNs is not robust for transformative features\cite{sainath2013deep}. Some new researchs find convolutional neural networks (CNNs) have strong abilities to explore inherit and hidden patterns through huge amount of training data. Several attempts that apply CNN to ESC have received performance boosts by learning spectrogram-like features from environment sounds\cite{Zhang2015Robust, Piczak2015Environmental, Salamon2016Deep}. However, the existing networks for ESC mostly use shallow architecture, such as 2 convolutional layers\cite{Piczak2015Environmental,Zhang2015Robust} and 3 convolutional layers\cite{Salamon2016Deep}. Getting a more discriminative and powerful information usually requests a deeper model. Therefore in this paper, we propose an enhanced CNN architecture with a deeper network based on VGG Net\cite{Simonyan2014Very}. The main contributions of this paper includes
\begin{itemize}
    \item  We propose a novel CNN network based on VGG Net. We find that simply using stacked convolutional layers with 3x3 convolution filters is unsatisfying in our tasks. So we redesign a novel CNN architecture in our ESC system. Instead of 3x3 convolution filters, We use 1-D convolution filters to learn local patterns across frequency and time, respectively. And our method performs better than CNN using 3x3 convolution filters with same depth of network.  
    \item  Mixup is applied in our ESC system for ESC tasks. Every training sample is created by mixing two examples randomly selected from original training dataset when using mixup. And the training target is also changed to the mix ratio. The effectiveness of mixup on classification performance and feature distribution is then explored further. 
    \item  Experiments were conducted on UrbanSound8K, ESC-50 and ESC-10 datasets, the result of which demonstrated that our ESC system has achieved the state-of-the-art performance (83.7$\%$) on UrbanSound8K and competitive performance on ESC-50 and ESC-10. 
\end{itemize}

The rest of this paper is organized as follows. Recent related works of ESC are introduced in Section~\ref{sect:work}. Section~\ref{sect:sys} provides detailed introduction of our methods. Section~\ref{sect:exp} presents the experiments settings on ESC-10, ESC-50 and UrbanSound8K datasets, and Section~\ref{sect:result and analysis} gives both experimental results and detailed discussions of our results. Finally, Section~\ref{sect:conc} concludes the paper.

\section{Related Work}
\label{sect:work}
In this section, we introduce the recent deep learning methods for environmental sound classification. Piczak\cite{Piczak2015ESC} proposed to apply CNNs to the log mel spectrogram which is calculated for each frame of audio and represents the squared magnitude of each frequency area. Piczak created a two-channel feature by applying log mel spectrogram and its delta information as the input of his CNN model and gave a 20.5$\%$ improvement over Random Forest method on ESC-50 dataset. Takahashi et al.\cite{takahashi2016deep} also used log mel spectrogram and their delta and delta-delta information as a three-channel input in a manner similar to the RGB inputs of the image. Dharmesh et al.\cite{agrawal2017novel} used gammatone spectrogram and a similar CNN architecture as Piczak [18] and claimed that they achieved 79.1$\%$ and 85.34$\%$ accuracy on ESC-50 and UrbanSound8K dataset, respectively. However, since their results were not reproducible, we contacted with the author and realized that the results achieved by them didn't follow the official cross validation methods, which means they used different training data and validation data than main published papers and not comparable.  So we will not compare our results with the results from \cite{agrawal2017novel}.

Some researchers also proposed to train model directly from raw waveforms. Dai et al.\cite{dai2017very} proposed a deep CNN architecture (up to 34 layers) with 1-D convolutional layers using 1-D raw data as input and they showed competitive accuracy with CNN using log mel spectrogram inputs\cite{Piczak2015ESC}. Tokozume et al.\cite{tokozume2017learning} proposed a end-to-end network named EnvNet using raw data as inputs and reported EnvNet could extract a discriminative feature that complements the log mel features. In \cite{tokozume2018learning}, they constructed a deeper recognition network based on EnvNet, referred as EnvNet-v2, and achieved better performance.

In addition, some researchers proposed to use external data for sound recognition. Mun et al.\cite{mun2017deep} proposed a DNN based transfer learning method for ESC. They first trained a DNN model using merged different web accessible environmental sound datasets. Then, they transferred the parameters of the pre-trained model and adapted the sound recognition system for target domain task using additional layers. Aytar et al.\cite{aytar2016soundnet} proposed to learn rich sound representations from large amounts of unlabeled sound and videos dataset. They transferred the knowledge of pre-trained visual recognition network into the sound recognition network. Then, they used a linear-SVM classifier to classify the feature which is the output of the hidden layer of the sound recognition network to the target task.

\section{Methods} \label{sect:sys}
\subsection{Convolutional Neural Network}
CNN is a stack of multi-layer neural networks including a group of convolutional layers, pooling layers and a limited number of fully connected layers. 
In this section, we propose a novel CNN as our ESC system model inspired by VGG Net\cite{Simonyan2014Very}, the architecture of which is presented in Table 1.
The proposed CNN architecture is comprised of eight convolutional layers and two fully connected layers. We first use 2 convolutional layers with large filter kernals as a basic feature extractor. Then, we learn local patterns across frequency and time using 3x1 and 1x5 convolution filters, respectively. Next, we use small convolution filters (3x3) to learn joint time-frequency patterns. Batch normalization\cite{Ioffe2015Batch} is applied to the output of convolutional layers to speed up training. We use the Rectified Linear Units (ReLU) to model the non-linearly for the output of each layer. After every two convolutional layers, a pooling layer is used to reduce the dimensions of the convolutional features maps, where maximum pooling is chosen in our network. To reduce the risks of overfitting, the dropout technique is applied after the first fully connected layers, with the probability of $0.5$. L2-regularization is applied to the weights of each layer with the coefficient $0.0001$. In the output layer, softmax function is used as the activation function which outputs probabilities of all classes.

\begin{table}[h]
\caption{Configuration of proposed CNN. Out shape represents the dimension in (frequency, time, channel). Batch Normalization is applied for each convolutional layer.
\label{CNN}}
\centering  
\setlength{\tabcolsep}{2mm}{
\begin{tabular}{lcccccccc}
\hline
Layer &Ksize &Stride &Nums of filters &Out shape\\
\hline
Input &- &- &- &(128, 128, 2)\\
\hline
Conv1 &(3, 7) &(1, 1) &32 &(128, 128, 32)\\
Conv2 &(3, 5) &(1, 1) &32 &(128, 128, 32)\\
Pool1 &(4, 3) &(4, 3) &- &(32, 43, 32)\\
\hline
Conv3 &(3, 1) &(1, 1) &64 &(32, 43, 64)\\
Conv4 &(3, 1) &(1, 1) &64 &(32, 43, 64)\\
Pool2 &(4, 1) &(4, 1) &- &(8, 43, 64)\\
\hline
Conv5 &(1, 5) &(1, 1) &128 &(8, 43, 128)\\
Conv6 &(1, 5) &(1, 1) &128 &(8, 43, 128)\\
Pool3 &(1, 3) &(1, 3) &- &(8, 15, 128)\\
\hline
Conv7 &(3, 3) &(1, 1) &256 &(8, 15, 256)\\
Conv8 &(3, 3) &(1, 1) &256 &(8, 15, 256)\\
Pool4 &(2, 2) &(2, 2) &- &(4, 8, 256)\\
\hline
FC1 &- &- &512 &(512, )\\
FC2 &- &- &nums of classes &(nums of classes, )\\
\hline
\end{tabular}}
\end{table}

\subsection{Mixup}
\label{subsect:mixup}

\begin{figure}
\centering
        \includegraphics[width=4.8 in]{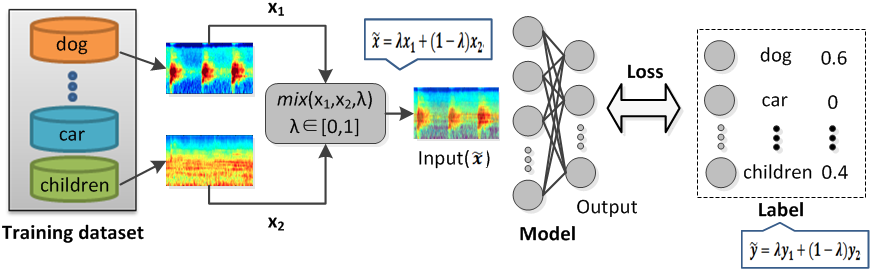}
        \caption{Pipeline of mixup. Every training sample is created by mixing two examples randomly selected from original training dataset. We use the mixed sound to train the model and the train target is the mixing ratio.}
        \label{fig:mixup}
\end{figure}

Mixup is an simple but effective method to generate training data\cite{zhang2017mixup}. Fig \ref{fig:mixup} shows the pipeline of mixup. Different from traditional augmentation approaches, mixup constructs virtual training samples by mixing training samples. Normally, a model is optimized by using a mini-batch optimization method, such as mini-batch SGD, and each mini-batch data is selected from the whole original training data. In mixup, however, each data and label of a mini-batch is generated by mixing two training samples, which are determined by 
\begin{equation}\label{eqn:mapping}
\left\{
\begin{aligned}
\hat{\mathbf x}= {\lambda}x_i + (1-\lambda)x_j\\
\hat{\mathbf y}= {\lambda}y_i + (1-\lambda)y_j
\end{aligned}
\right.
\end{equation}
where $x_i$ and $x_j$ are two samples randomly selected from training data, and $y_i$ and $y_j$ are their one-hot labels. The mix factor $\lambda$ is decided by a hyper-parameter $\alpha$ and $\lambda$ $\sim$ Beta($\alpha$, $\alpha$). Therefore, mixup extends the training data distribution by mixing various training data within or without the same class by a linear way, leading to a linear interpolation of the associated targets. Note that we do not use mixup for testing phase.

\section{Experiments} \label{sect:exp}

\subsection{Dataset}

Three publicly available datasets are used for model training and performance evaluation of the proposed approach, including ESC-10, ESC-50\cite{Piczak2015ESC} and UrbanSound8K\cite{salamon2014dataset}, the detailed information of which is shown in Table \ref{tab:dataset}.

The ESC-50 dataset consists of 2000 short environmental records which are divided into 50 classes in 5 major categories, including \emph{animals, natural soundscapes and water sounds, human non-speech sounds, interior/domestic sounds}, and \emph{exterior/urban noises}. All audio samples are 5 seconds with 44.1 kHz sampling frequency.

The ESC-10 dataset is a subset of 10 classes (400 samples) selected from the ESC-50 dataset (\emph{dog bark, rain, sea waves, baby cry, clock tick, person sneeze, helicopter, chainsaw, rooster, fire crackling}).

The UrbanSound8K dataset is a collection of 8732 short (up to 4 seconds) audio clips of urban sound areas. And the audio clips are prearranged into 10 folds. The dataset is divided into 10 classes: \emph{air conditioner, car horn, children playing, dog bark, drilling, engine idling, gun shot, jackhammer, siren}, and \emph{street music}.

\begin{table}[t]
\caption{Information of datasets.
\label{tab:dataset}}
\centering  
\setlength{\tabcolsep}{2mm}{
\begin{tabular}{cccc}
\hline
Datasets &Classes &Nums of samples &Duration\\
\hline
UrbanSound8K &10 &8732 &9.7 hours \\
\hline
ESC-50 &50 &2000 &2.8 hours \\
\hline
ESC-10 &10 &400 &33 min \\
\hline
\end{tabular}}
\end{table}

\subsection{Preprocessing}
We use a 44.1kHz sampling rate for ESC-10, ESC-50, UrbanSound8K datasets. 
All audio samples are normalized into a range from $-1$ to $1$. In order to avoid overfitting and to effectively utilize the limited data, we use Time Stretch\cite{Salamon2016Deep} and Pitch Shift\cite{Salamon2016Deep} deformation methods to generate new audio samples. We use two spectrogram-like representations, log mel spectrogram (Mels) and gammatone spectrogram (GTs). Both features are extracted from all recordings with hamming window size of 1024, hop length of 512 and 128 bands. Then, the resulting spectrograms are converted into logarithmic scale. In our experiments, we use a simple energy-based silence drop algorithm to drop silence regions. Finally, the spectrograms are split into 128 frames (approximately $1.5s$) length with 50$\%$ overlap. The delta information of the original spectrogram is calculated, which is the first temporal derivative of the spectrogram feature. Then, we use the segments with their deltas as a two-channel input to the network. 

\subsection{Training settings}
All models are trained using mini-batch stochastic gradient descent (SGD) with Nesterov momentum of 0.9. We used a learning rate decrease schedule with a initial learning rate of 0.1, and then divided the learning rate by 10 every 80 epoch for UrbanSound8K and 100 epoch for ESC-10 and ESC-50. Every batch consists of 200 samples randomly selected from training set without repetition. The models are trained for 200 epochs for UrbanSound8K and 300 epochs for ESC-50 and ESC-10. We initialize all the weights to zero mean Gaussian noise with a standard deviation of 0.05. We use cross entropy as the loss function, which is typically used for multi classification task.

In the test stage, feature extraction and audio cropping patterns are the same as those used in the training stage. Prediction probability of a test audio sample is the average of predicted class probability of each segment. The predicted label of the test audio sample is the class with the highest posterior possibility. The classification performance of the methods is evaluated by the $K$-fold cross-validation. For the ESC-50 and ESC-10 dataset, $K$ is set to $5$, while for the UrbanSound8K dataset, $K$ is set to $10$. 

All models are trained using Keras library with TensorFlow backend on an Nvidia P100 GPU with a 12GB memory.

\section{Results and Analysis}
\label{sect:result and analysis}

The classification accuracy of the proposed method compared with recent related works is shown in Table \ref{tab:accuracy}. It can be observed that our method achieved the state-of-the-art performance (83.7$\%$) on UrbanSound8K dataset and competitive performance (91.7$\%$, 83.9$\%$) on ESC-10 and ESC-50. 
The average classification accuracy of our methods with Mels outperformed PiczakCNN\cite{Piczak2015Environmental} (baseline) by 10.8$\%$, 17.6$\%$, 9.9$\%$ on ESC-10, ESC-50 and UrbanSound8K datasets, respectively. 
Data augmentation is an important technique for increasing performance for limited dataset, which gave an improvement of 1.1$\%$, 3.3$\%$ and 5.3$\%$ on ESC-10, ESC-50 and UrbanSound8K, respectively. 
In addition, GTs improved by 0.4$\%$, 1.4$\%$ and 1.1$\%$ over Mels on ESC-10, ESC-50 and UrbanSound8K, respectively. We can see that classification accuracy with GTs is always better than accuracy with Mels on on ESC-10, ESC-50 and UrbanSound8K datasets, which indicates that feature representation is a critical factor for classification performance. What's more, mixup is a powerful way to improve performance which can always perform better results than that without mixup. In our experiments, Mixup gave an improvement of 1.5$\%$, 2.4$\%$ and 2.6$\%$ with Mels on ESC-10, ESC-50, UrbanSound8k datasets, respectively. As mentioned in Section \ref{sect:sys}, mixup trains a network using a linear combination of training examples and their labels and leads to a regularization for neural network and generalization for unseen data. For the effect of mixup, we do a further exploration in the following parts.

\begin{table}[tb]
\caption{Classification accuracy ($\%$) of different ESC systems. In our ESC system, we compare two different features with augmentation and without augmentation. 'aug' stands for augmentation, including Pitch Shift, Time Stretch. Note that we will not compare with the results of Dharmesh\cite{agrawal2017novel} which was discussed in Section \ref{sect:work}. \label{tab:accuracy}}
\centering  
\setlength{\tabcolsep}{2mm}{
\begin{tabular}{ccccc}
\hline
\multicolumn{2}{c}{} &\multicolumn{3}{c}{Acc ($\%$)}\\
\hline
Model &Feature &ESC10 &ESC50 &UrbanSound8K\\
\hline
PiczakCNN\cite{Piczak2015Environmental} &Mels &80.5 &64.9 &72.7\\
\hline
D-CNN\cite{zhang2017dilated} &Mels &- &68.1 &81.9\\
\hline
SoundNet\cite{aytar2016soundnet} &- &{\bfseries 92.1} &74.2 &-\\
\hline
Envnet-v2\cite{tokozume2017learning} &Raw data &91.4 &{\bfseries 84.9} &78.3\\
\hline
\multirow{2}*{proposedCNN} &Mels &88.7 &76.8 &74.7\\
~ &GTs &89.2 &78.9 &77.4\\
\hline
\multirow{2}*{proposedCNN + mixup} &Mels &90.2 &79.2 &77.3\\
~ &GTs &90.7 &80.7 &79.8\\
\hline
\multirow{2}*{proposedCNN + aug + mixup} &Mels &91.3 &82.5 &82.6\\
~ &GTs &{\bfseries 91.7} &83.9 &{\bfseries 83.7}\\
\hline
\hline
human performance &- &95.7 &81.3 &-\\
\hline
\hline
Dharmesh\cite{agrawal2017novel} &GTs &- &79.10 &85.34\\
\hline
\end{tabular}}
\end{table}

\begin{figure}[t]
  \centering
  \subfigure[]{
    \includegraphics[width=2.2in]{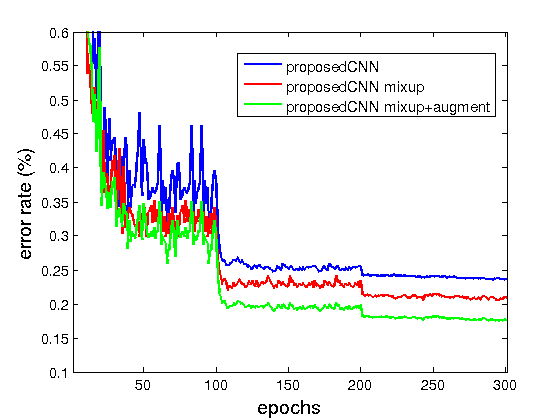}
  }
  \subfigure[]{
    \includegraphics[width=2.2in]{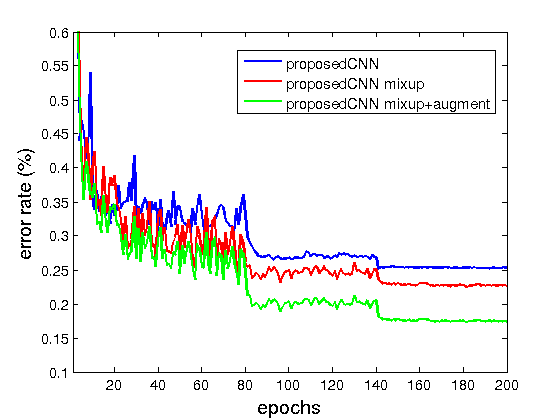}
  }
  \caption{Training curves of our proposed CNN on (a) ESC-50 and (b) UrbanSound8K datasets.} 
\end{figure}

\subsection{Comparison of network architecture}

We compare our proposed CNN with a VGG network architecture with same depth of network. This VGG network has same network parameters with our proposed CNN except for replacing to use 3x3 convolution filters and 2x2 stride pooling and we refer to this architecture as VGG10. In Table \ref{tab:cnet}, we provide classification accuracy of proposedCNN and VGG10 on ESC-10, ESC-50 and UrbanSound8K datasets. The results shows that our proposed CNN always performs better than VGG10 on three datasets. 

\begin{table}[t]
\caption{Comparison between proposed CNN and VGG10 Net ($\%$).
\label{tab:cnet}}
\centering  
\setlength{\tabcolsep}{2mm}{
\begin{tabular}{cccc}
\hline
Model &ESC-10 &ESC-50 &UrbanSound8K\\
\hline
proposedCNN &88.7 &76.8 &74.7 \\
\hline
VGG10 &87.5 &73.3 &73.2 \\
\hline
\end{tabular}}
\end{table}

\subsection{Effects of Mixup}
\begin{figure}
  \centering
  \subfigure[]{
    \includegraphics[width=2.2in]{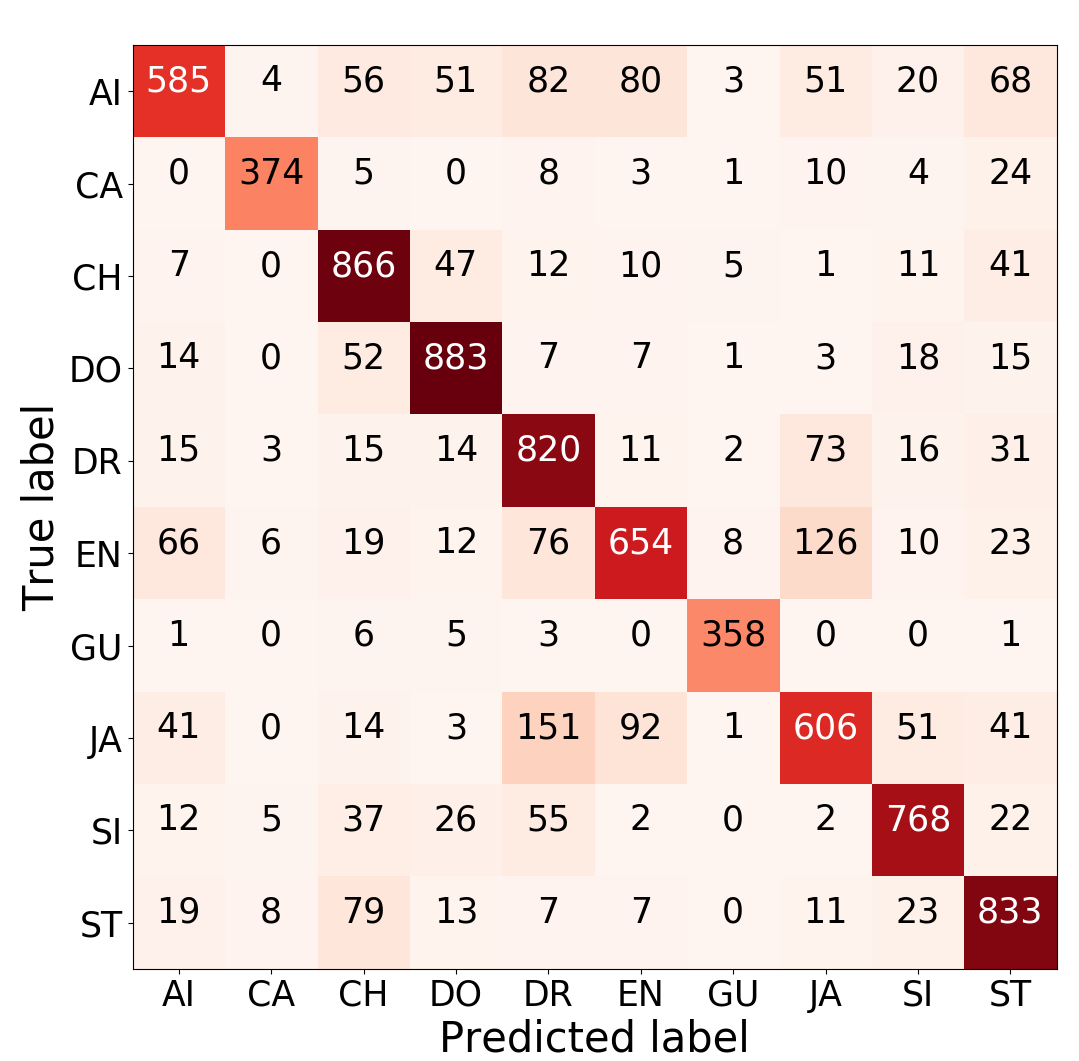}
  }
  \subfigure[]{
    \includegraphics[width=2.2in]{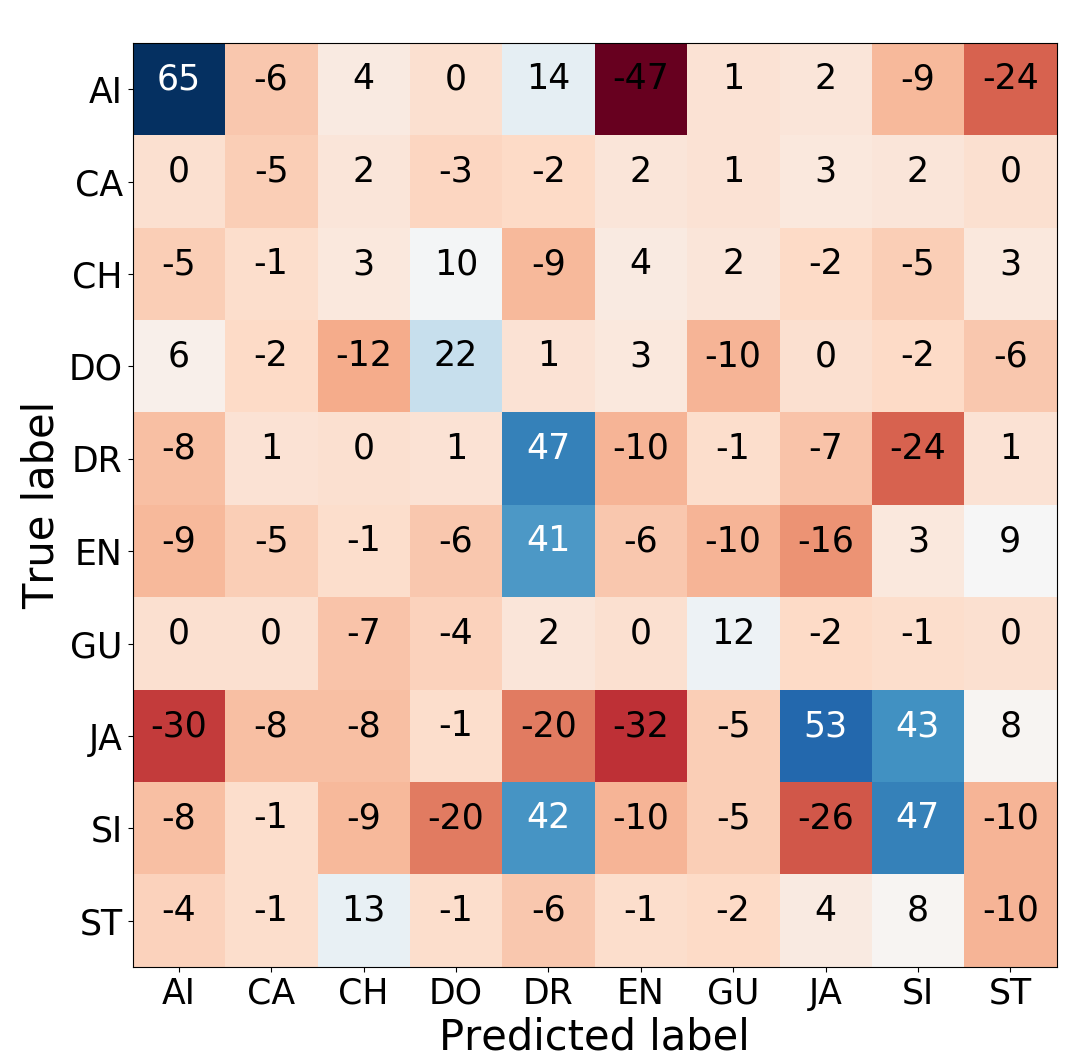}
  }
  \caption{(a) Confusion matrix for UrbanSound8K dataset using the proposed CNN model applying to Mels with mixup augmentation methods. (b) Different between the confusion matrix for UrbanSound8K dataset using the proposed CNN and Mels with mixup and without mixup: the negative values (brown) mean the confusion is decreased with mixup, the positive (blue) values mean the confusion is increased with mixup. Classes are air conditioner (AI), car horn (CA), children playing (CH), dog barking (DO), drilling (DR), engine idling (EN), gun shot (GU), jackhammer (JA), siren (SI) and street music (ST).}
  \label{fig:matrix} 
\end{figure}

{\bfseries Analysis.}  
The confusion matrix by the proposed CNN with Mels and mixup for the UrbanSound8K dataset is given in Fig.\ref{fig:matrix} (a).
We can observe that the most misrecognition happened between two noise-like classes, such as \emph{jackhammer} and \emph{drilling}, \emph{engine idling} and \emph{jackhammer}, and \emph{air conditioner} and \emph{engine idling}. 
In Fig.\ref{fig:matrix} (b), we provide the difference of the confusion for the proposed CNN method with and without mixup. 
We see that mixup gives an improvement for most classes, especially for \emph{air conditioner, drilling, jackhammer} and \emph{siren}. 
However, mixup also has a slightly harmful effect on the accuracy for some classes and increases confusion between some specific pairs classes. For example, although mixup reduces the confusion between \emph{jackhammer} and \emph{engine idling}, it increases the confusion between \emph{jackhammer} and \emph{siren}.

\begin{figure}
  \centering
  \subfigure[]{
    \includegraphics[width=2.2in]{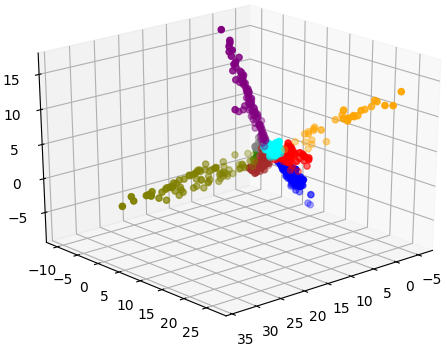}
  }
  \subfigure[]{
    \includegraphics[width=2.2in]{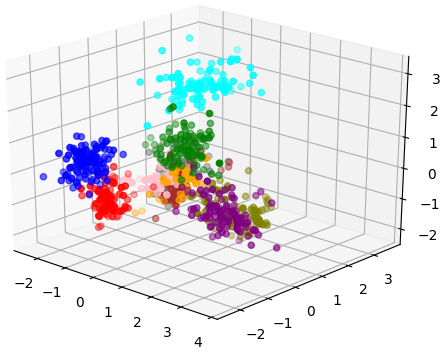}
  }
  \caption{Visualization of the feature distribution at the output of FC1 using PCA (a) without mixup and (b) with mixup.} 
  \label{fig:visualization} 
\end{figure}

To gain further insights to the effect of mixup, we visualized the feature distributions for UrbanSound8K with mixup and without mixup using PCA in Fig.\ref{fig:visualization}. The feature dots represent the high-level feature vectors obtained at the output of the first fully connected layer (FC1). We can observe that it is quite different between feature distributions with and without mixup. Fig.\ref{fig:visualization} (a) shows the feature distributions of different classes with mixup. Some classes have a large within-class variance of the feature distribution, while some have a small within-class variance. In addition, the between-class distances of different pairs of classes are also varied, which may make models more sensitive to some classes. However, features of most classes distribute within a small space with a relative smaller within-class variance and the boundary of most classes is clear as shown in Fig.\ref{fig:visualization}(b).

\begin{figure}[tb]
\centering
        \includegraphics[width=3 in]{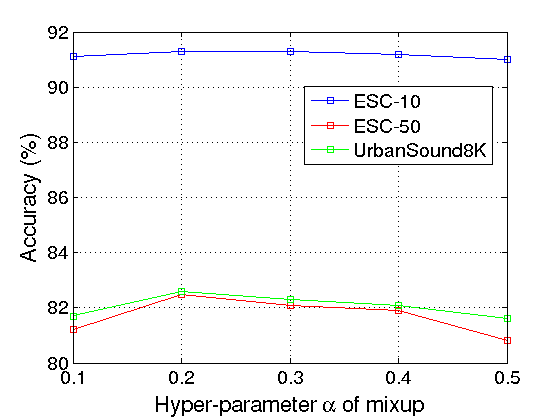}
        \caption{Curves of an accuracy with different $\alpha$ for ESC-10, ESC-50, UrbanSound8K}
        \label{fig:para_curve}
\end{figure}

\noindent {\bfseries Hyper-parameter $\alpha$ selected.}  In order to achieve a better performance for our system on ESC, the effect of mixup hyper-parameter $\alpha$ is further explored. Fig.\ref{fig:para_curve} shows the change of accuracy with different $\alpha$ ranging from $[0.1, 0.5]$. We see that when $\alpha = 0.2$, the best accuracy is achieved on all three datasets. 

\section{Conclusion} \label{sect:conc}
In this paper, we proposed a novel deep convolutional neural network architecture for environmental sound classification. We compared our proposed CNN with VGG10 and results showed that our proposed CNN always performed better. To further improve the classification accuracy, mixup was applied in our ESC system. As a result, the proposed ESC system achieved state-of-the-art performance on UrbanSound8K dataset and competitive performance on ESC-10 and ESC-50 dataset. Furthermore, we explored the impacts of mixup on the classification accuracy and feature space distribution of different classes on UrbanSound8K dataset. The results showed that mixup is a powerful method to improves classification accuracy. Our future work will focus on the network design and exploration for using mixup method for specific classes.

\bibliographystyle{splncs04}
\bibliography{myreference}
\end{document}